# Robust Audio Watermarking Algorithm Based on Moving Average and DCT

Jinquan Zhang and Bin Han
(*College of Information security Engineering,*
*Chengdu University of Information Technology, Chengdu, China*)
*zhjq@cuit.edu.cn*，hanbin@ *cuit.edu.cn*

*Abstract*

*Noise is often brought to host audio by common signal processing operation, and it usually changes the high-frequency component of an audio signal. So embedding watermark by adjusting low-frequency coefficient can improve the robustness of a watermark scheme. Moving Average sequence is a low-frequency feature of an audio signal. This work proposed a method which embedding watermark into the maximal coefficient in discrete cosine transform domain of a moving average sequence. Subjective and objective tests reveal that the proposed watermarking scheme maintains highly audio quality, and simultaneously, the algorithm is highly robust to common digital signal processing operations, including additive noise, sampling rate change, bit resolution transformation, MP3 compression, and random cropping, especially low-pass filtering.*

*Keywords: audio watermarking; robust watermarking; moving average sequence; discrete cosine transform*

## 1. Introduction

With the development of Internet network, more and more resources are fully shared on Internet. But some copyrighted digital products are also spread in unauthorized circumstances, which seriously harmed the creators' interests, and hurt the author's creative passion. An important method to track piracy is to embed robust digital watermark in the works.

Audio watermarking has been attracting the researchers' attention [1]. Some audio watermarking algorithms embedded information in a single domain. For examples, literature [2-4] embedded information in time domain, and literature [5-7] embedded watermark in frequency domain. Usually, algorithm which embeds information in time domain has high efficiency, and algorithm which embeds information in the frequency domain has good robustness to a variety of signal processing operations.

For signal processing transformations have respective advantages, some researchers combined good properties of two or three kinds of transformations and embedded watermark into hybrid domain. Making use of the multi-resolution of discrete wavelet transform(DWT) and the energy compression of discrete cosine transform domain(DCT), Wang and Zhao [8] argued embedding the watermark into the hybrid domain. Literature [9] proposed an improved algorithm based on literature [3], which combined characteristics of DWT and histogram. The robustness to MP3 compression and low-pass filter was increased. Literature [10] performed DWT on a frame audio, then, singular value decomposition(SVD) was done on the approximation coefficients, and embedded watermark into the SVD coefficients. Literature [11] segmented audio signal and SVD was performed on each segment, then first SVD coefficient in every block were grouped together and DCT was done, and the watermark was embedded into DCT coefficients. These algorithms all had good robustness.

To a robust audio watermarking algorithm, the watermark information is required to survive common signal processing operations, such as MP3 compression, low-pass filter and so on. As we known, the time delay will take place when low-pass filtering is done on a watermarked audio, and about 1000 zero samples will be added at the front of the audio signal when MP3 compression is performed on. At the same time, the random-clipping may be happen to a watermarked audio. It is necessary to embed synchronization code into a watermarked audio [2, 10-12].

For common signal operations mainly change the high frequency component of a signal, embedding message in low frequency component of an audio signal is important to improve the robustness of an algorithm. In literature [13-17], DWT was performed on audio signal segment, then, the message was embedded in approximate coefficient. In literature [8, 10, 18], DWT was performed the audio signal frame first, and other transformation are carried out on the approximate coefficient, then, the message was embedded in the coefficient of hybrid domain.

Decreasing the embedding capacity is also a way to improve the robustness, for decreasing the capacity means that embedding a bit message involves more samples, the characteristic of the signal segment is more stable. In literature [2], embedding one bit message needs 1020 samples. If the sampling frequency of an audio signal is 44.1kHz, the embedding capacity of the algorithm is about 43bps. In literature [3], the value is 2bps. In literature [19], it is 80bps and in literature [11], it is 43bps.

In our scheme, a moving average sequence(MAS) of audio signal is obtained first. It is the low-frequency feature of an audio signal. Then, DCT is performed on a long sequence segment to get a stable feature. The maximal coefficient of DCT is chosen to embed message by QIM [20]. At the same time, synchronization code is used to locate the watermark and improve the accuracy of the detecting algorithm.

The rest of this paper is organized as follows: In section 2, we defines the moving average sequence, describes its properties, and presents an rapid calculation of MAS. Section 3 separately describes the watermark embedding and watermark extraction of the proposed watermarking method. In section 4, the experimental results are presented to show the performance of the proposed watermark. The conclusion of this paper is drawn out in section 5.

## 2. Definition of MAS and its property

To explain our algorithm conveniently, we give the definition of moving average sequence here, and describe some properties of it below.

### 2.1 Definition of MAS and its property

Assume the sample number of an audio clip is R, and the value of them are denoted as $x_1, x_2, \ldots, x_R$. Choose an integer $b$, and the MAS $M_B$ is defined as equation (1).

$$M_{B_i} = \frac{1}{b}(x_i + x_{i+1} + \cdots + x_{i+b-1}) = \frac{1}{b}\sum_{k=i}^{i+b-1} x_k, i \in (1, R-b+1) \qquad (1)$$

It is easy to know from the equation (1) that the sequence has good low-pass characteristic. As we all know, common signal processing, such as additive noise, sampling rate change, bit resolution transformation and so on, usually mainly changes high-frequency component of an audio signal, but the low-frequency component is slightly distorted.

Our experimental results shows that, for an audio clip, the MAS is nearly unchanged after it undergo common signal processing operations, such as additive noise, sampling rate change, bit resolution transformation, and so on, even the audio waveform has obviously different. That is, their MASs are nearly overlapping, which means their low-frequency components are almost the same.

In the following experiment, the low-pass filter operation is done to a certain audio clip. In the experiment, the integer $b$=30, the cutoff frequency of low-pass filter is 8kHz. The waveform of the original audio clip and its MAS, the waveform of audio clip after low-pass filtering and its MAS are all shown in figure 1. As we all know, the time delay will take place when low-pass filter is done to a time signal. In order to compare the shapes of these waveforms, the waveform after low-pass filter processing and its MAS are both shift to left 3 sampling periods. As can be seen from figure 1, the waveform of the audio clip changed obviously after low-pass filtering, but their MASs are almost overlapping.

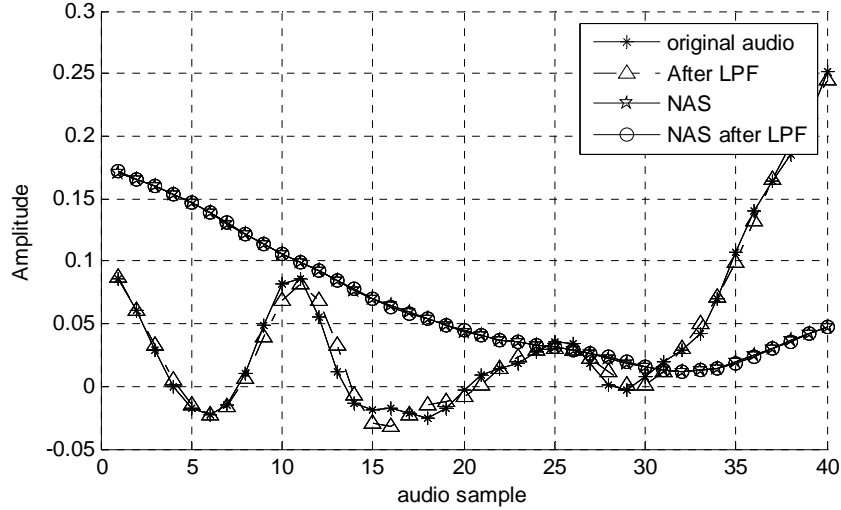

**Figure 1. audio waveform and its MAS before and after low-pass filtering**

### 2.2 Rapid calculation of MAS

The following method is used to obtain the MAS of an audio segment quickly.

Assume an audio signal X=($x_1$, $x_2$, …, $x_R$). For a given integer b>0,

Let
$$M_{B_i} = (x_i + x_{i+1} + \cdots + x_{i+b-1})/b \quad (2)$$
Now, we will compute $M_{B_{i+1}}$. In fact, compare with $M_{B_i}$, when the $M_{B_{i+1}}$ is computed, the $i$th sample is excluded, and the ($i+b$)th sample is included.
Then
$$M_{B_{i+1}} = M_{B_i} + x_{i+b}/b - x_i/b \quad (3)$$
From equation (3), we can obtain equation (4) as follows.
$$M_{B_{i+1}} = M_{B_i} + (x_{i+b} - x_i)/b \quad (4)$$
The method will reduce the computation load dramatically.

## 3. Embedding and extracting watermark

For the MAS shape of an audio clip is almost the same before and after signal processing operations, the proposed scheme embeds message in DCT domain of a MAS. The scheme of the synchronization code described in literature [21] is adopted to locate the watermark information. If the audio clip is very long, the synchronization code and watermark are embedded repeatedly.

### 3.1 Embedding watermark

We choose an integer *b* to calculate the MAS $M_B$ following equation (1) according to our definition in section 2. In the rest of this paper, for convenient writing, we substitute *M* for $M_B$. Similarly, we substitute $M_i$ for $M_{B_i}$. That is to say, *b* samples, $x_i$, $x_{i+1}$, …, $x_{i+b-1}$, are involved to compute the $M_i$ according to equation (1).

We assume the length of the watermark message *w* is *L*, and the *w* is a pseudo-random sequence made up of {-1,1}. In order to acquire the *w*, first, we may obtain a pseudo-random sequence composed of 0 and 1, which can be generated from a linear feedback shift register, or a cryptographic algorithm standard, such as ANSI X9.32 and so on. Then, 0 is replaced with -1 in the sequence.

In the proposed scheme, after the synchronization code is embedded, the watermark information *w* will be embedded into the MAS.

After the synchronization code, one frame *M* was chosen to embed into one bit message. The length of each frame is $n \times b$. That is to say, in order to embed the message *w*, the length $n \times b \times L$ of *M* is needed. How to choose the parameters n and b is presented in section 4.1.

For the 1-th frame of M after the synchronization code, we denoted it as ($M_1$, $M_2$, …, $M_{n \times b}$), which involves $n \times b + b - 1$ samples, denoted as ($x_1$, $x_2$, …, $x_{n \times b + b - 1}$). Then, For a certain frame of *M*, such as *i*-th frame, it is denoted as ($M_{n \times b \times i + 1}$, $M_{n \times b \times i + 2}$, …, $M_{n \times b \times (i+1)}$), which involves $n \times b + b - 1$ samples, denoted as ($x_{n \times b \times i + 1}$, $x_{n \times b \times i + 2}$, …, $x_{n \times b \times (i+1) + b - 1}$).

These expressions in the previous paragraph weren't simplified because it is helpful to understand our scheme, such as $n \times b \times i + b + 1$, $n \times b \times i + n \times b$ and so on.

In our scheme, for avoiding the interference which is brought by the previous bit message embedding, the sequence ($M_{n \times b \times i + 1}$, $M_{n \times b \times i + 2}$, …, $M_{n \times b \times i + b}$) keeps unchanged, and DCT is performed on ($M_{n \times b \times i + b + 1}$, $M_{n \times b \times i + b + 2}$, …, $M_{n \times b \times i + n \times b}$). Then, the QIM[20] is adopted to embed watermark message by modifying the maximal coefficient of DCT. That is to say, the length of DCT is $(n-1) \times b$.

For the *i*-th frame of $M_B$, the detailed procedure of embedding one bit message into it is described as follows.

(1) Set the embedding strength of the scheme $S > 0$. Let $d[-1] = 3S/4$, $d[1] = S/4$.

(2) If the number of the remaining audio samples may embed complete watermark message *w* one time, go to step(3), if not, the embedding procedure is end.

(3) DCT is performed on ($M_{n \times b \times i + b + 1}$, $M_{n \times b \times i + b + 2}$, …, $M_{n \times b \times i + n \times b}$). Assume the coefficient with the largest absolute value is *t* in DCT coefficients. So, *t* is a signed number. According to the properties of DCT, it must be a low frequency coefficient and relatively stable to common signal processing operation. Then, the *i*-th bit watermark information *w(k)* will be embedded. The embedding rule is shown in equation(5).

$$t' = \begin{cases} round((t + d[1])/S) \times S - d[1], & \text{if } w(i) = 1 \\ round((t + d[-1])/S) \times S - d[-1], & \text{if } w(i) = -1 \end{cases} \quad (5)$$

Where round(•) means rounding to nearest integer.

(4) The *t'* may not be the maximal one now, so we must adjust coefficients whose absolute value is larger than |*t'*|. To improve the robustness of the scheme, coefficients whose absolute value is close to |*t'*| are also slightly reduced. An optional approach is that, if the |*t'*|>S, these coefficients are changed to |*t'*|-S/8, if not, changed to 0.8S. The sign of these coefficients is unchanged, and |•| indicates the absolute value function. Other adjustment ways may be also acceptable.

(5) IDCT is performed and ($M'_{n \times b \times i + b + 1}$, $M'_{n \times b \times i + b + 2}$, …, $M'_{n \times b \times i + n*b}$) is obtained. Then, subtract ($M_{n \times b \times i + b + 1}$, $M_{n \times b \times i + b + 2}$, …, $M_{n \times b \times i + n \times b}$) from ($M'_{n \times b \times i + b + 1}$, $M'_{n \times b \times i + b + 2}$, …, $M'_{n \times b \times i + n*b}$) correspondingly. Set the variation of sequence ($M_{n \times b \times i + 1}$, $M_{n \times b \times i + 2}$, …, $M_{n \times b \times i + n*b}$) is *v*, then

$v = (M_{n \times b \times i + 1} - M_{n \times b \times i + 1}, M_{n \times b \times i + 2} - M_{n \times b \times i + 2}, …, M_{n \times b \times i + b} - M_{n \times b \times i + b},$
$M'_{n \times b \times i + b + 1} - M_{n \times b \times i + b + 1}, M'_{n \times b \times i + b + 2} - M_{n \times b \times i + b + 2}, …, M'_{n \times b \times i + n*b} - M_{n \times b \times i + n \times b})$
$= (v_1, …, v_b, v_{b+1}, …, v_{n \times b})$
$= (0, …, 0, v_{b+1}, …, v_{n \times b}).$

There are *b* zeros in the front of the sequence because ($M_{n \times b \times i + 1}$, …, $M_{n \times b \times i + b}$) isn't involved into DCT. So they keep the same.

The variation of corresponding audio samples should satisfy the equation set:

$$\sum_{j=k}^{k+b-1} z_j = b \times v_k, k = 1, …n \times b \quad (6)$$

$z_j$ denotes variation of the ($n{\times}b{\times}i{+}j$)-th audio sample.

Now, we solve the equation set made up of the last $b$ equations. That is, the equation set is made up of equations from (($n{-}1$)$\times b{+}1$)-th to ($n{\times}b$)-th.

From the last equation, that is,

$$\sum_{j=n\times b}^{n\times b+b-1} z_j = bv_{n\times b} \tag{7}$$

We let $z_{n\times b+j} = v_{n\times b}$, $j = 0,1,2,...,b-1$.

Then, the other $b$-1 equations is solved,

$$z_{(n-1)\times b+j} = b(v_{(n-1)\times b+j} - v_{(n-1)\times b+j+1}) + z_{n\times b+j},\ j=1,2,...,b-1 \tag{8}$$

Then, the equation set, which is made up of equations from ($n$-2)$\times b$+1 to ($n$-1)$\times b$, is solved,

$$z_{(n-2)\times b+j} = b(v_{(n-2)\times b+j} - v_{(n-2)\times b+j+1}) + z_{(n-1)\times b+j},\ j=1,2,...,b \tag{9}$$

The rest equations can be solved in the same manner.

Assume the solution of equations (6) is $z_j$, $j \in [1, n{\times}b{+}b{-}1]$. Then, $z_j$, $j \in [1, n{\times}b{+}b{-}1]$ plus $x_r$, $r \in [n{\times}b{\times}i{+}1, n{\times}b{\times}i{+}n{\times}b{+}b{-}1]$ correspondingly, for example, $x'_{n\times b\times i+1} = x_{n\times b\times i+1} + z_1$. Then the watermarked audio is obtained.

### 3.2 Watermark extraction algorithm

Parameters which are used in embedding watermark phase are obtained first, including parameters used in the embed phase of synchronization code, the integer $b$, the length of frame $n{\times}b$, the embedding strength $S$, the length of the watermark $L$, and so on. The extraction scheme is described as follows.

The synchronization code is searched first. According to the synchronization code, the position where message is embedded into is located.

Similar to the embedding phase, After the synchronization code, the MAS $M''$(In fact, it is $M''_B$) is gotten by the watermarked audio according to equation (1) in section 2. Then, the MAS was chosen to detect message. The length of frame is also $n{\times}b$, and DCT is performed on each frame.

For a certain frame $M''_i$, set the maximal data of DCT coefficient is $t^*$, one bit message may be extracted as follows,

$$w_i = \begin{cases} 1, t^* - \lfloor t^*/S \rfloor \times S \geq S/2 \\ -1, t^* - \lfloor t^*/S \rfloor \times S < S/2 \end{cases} \tag{10}$$

The similar procedure is performed until all data are extracted.

Assume the number of watermark extraction is $n$, and the $i$-th message which is extracted from the watermarked audio is denoted as ($w^i_1$, $w^i_2$, ..., $w^i_L$). $L$ is the length of the watermark. Then, according to majority principle, the $k$-th bit watermark is obtained by the following equation.

$$w_k = \begin{cases} 1,\ if\ (\sum_{j=1}^{n} w^j_k) \geq 0 \\ -1,\ otherwise \end{cases} \tag{11}$$

So the watermark information is obtained.

## 4. Experimental results and discussions

In our experiments, a random sequence with 128 bits is adopted as watermark message, which is generated by the pseudo-random number algorithm in ANSI X9.32. The number of 1 and -1 are both 64 bits in the watermark. Many various styles of audio clips are chosen to accomplish the experiments. Each audio clip is a 16-bit mono audio clip in the WAVE format sampled at 44 100 Hz.

### 4.1 Choice of $b$ and $n$

Assume the number of zero-crossing of $M_{10}$ is $Z_{M_{10}}$, $M_{10}$ is obtained according to equation (1) in section 2. L is the sample number of the original audio clip. Let *num*=L/$Z_{M_{10}}$. The integer *b* in embedding phase in section 3 is less than *num* a little.

The parameter *n* is an experimental value. We choose many various styles of audio clips to accomplish the experiments. According to our experimental results, the algorithm have good compromise between robustness and audibility when n∈[8,10].

### 4.2 Imperceptibility test

In order to evaluate the quality of the watermarked audio, signal-to-noise ratio (SNR) and perceptual evaluation of audio quality (PEAQ), an assessment tool recommended by ITU BS1387, are both used in objective evaluation tests, and Mean Opinion Score (MOS), suggested by ITU-T P.800, is used in the subjective listening test. Here, we only report the results with three audio clips which are pop music, country music and blues music.

The parameters of the three chosen audio clips are as follows. For pop music clip, the length is 15 second, *b*=60, *n*=10, *S*=0.18. For country music clip, 15 second, *b*=60, *n*=10, *S*=0.18. For blues music clip, 16 second, *b*=50, *n*=9, *S*=0.2. For each clip, use PEAQ to measure the watermarked audio quality to adjust embedding parameters, especially the embedding strength *S*. In our experiments, the ODG(Objective Difference Grades) is less than -1, and the SNR is more than 30dB.

For those coefficients whose absolute value are larger than or close to |*t'*|, the way which is used to adjust these coefficients is as described in section 3.1. In our experiments, the largest coefficient is often the 3~12-th alternating current(AC) coefficient of the DCT domain.

In order to assess the algorithm, a comparative experiment which embed message in DCT is carried out. In the comparative experiment, the parameters and location of synchronization code are unchanging. The same message is embedded at the same location, and the embedding capacity is identical. The message is embedded into the maximal DCT coefficients by QIM rule. The embedding strength was adjusted according to the similar SNR.

The SNR and ODG of the watermarked audio are obtained in test. The subjective test is done by a team composed of 10 audiences. As shown in table 1, in the case of nearly SNR, our scheme have smaller ODG than the comparative scheme which message is embedded in the DCT domain directly, especially for soothing music. For example, pop music.

**Table 1 Imperceptibility test**

|  | Pop music | | Country music | | Blues music | |
| --- | --- | --- | --- | --- | --- | --- |
|  | Proposed scheme | DCT | Proposed scheme | DCT | Proposed scheme | DCT |
| SNR（dB） | 31.7 | 31.5 | 30.6 | 30.6 | 30.5 | 30.7 |
| ODG | -0.75 | -1.41 | -0.08 | -0.12 | -0.42 | -0.53 |
| MOS | 4.85 | 4.15 | 4.95 | 4.90 | 4.91 | 4.89 |

### 4.3 Embedding capacity

Assume the sampling frequency of the audio clip is *fs*, the embedding capacity is about *fs*/*nb* bits, *b* is the integer which is used to compute the MAS, and *n* is as described in section 2.1.

As mentioned above, the synchronization code is embedded before the message, as described in literature[21]. In proposed scheme, embedding one bit synchronization code involves about 2*b* samples. For one bit message, it is about 10*b* samples. At the same time, the length of the synchronization code is much less than the length of the watermark information. So the embedding capacity is mainly determined by the watermark embedding algorithm.

**4.4 Robustness tests**

In experiments, the processing operations shown in Table 2 are performed on the watermarked audio signals which are mentioned in section 3.1. These operations include:

(1) Additive white Gaussian noise (AWGN): white Gaussian noise is added to the watermarked signal until the resulting signal has an SNR of 55dB/45dB/35dB.

(2) Requantization: the 16-bit watermarked audio signals are re-quantized down to 8 bits/sample and then back to 16 bits/sample.

(3) Resampling: The watermarked signal, originally sampled at 44.1 kHz, is re-sampled at 22.05 kHz /11.025kHz, and then restored back by sampling again at 44.1 kHz.

(4) Low-pass filtering: A six order Butterworth filter with cut-off frequency 10kHz/8kHz/4kHz/2kHz is used.

(5) Cropping: Segments of one second are removed from the watermarked audio signal randomly.

(6) MP3 Compression 128kbps/96kbps/80 kbps/64 kbps: The MPEG-1 layer-3 compression is applied. The watermarked audio signal is compressed at the bit rate of 128kbps/96kbps and then decompressed back to the WAVE format.

The experimental results for above three audio clips are shown in Table 2. As can be seen in the table, the number of embedding message in the watermarked audio is 8, even the number of detecting watermark message is only 2-4, such as Additive noise (35dB), or MP3(64kbps), the bit error rate (BER) of the proposed algorithm is close to zero, or equal to zero, which means the algorithm has strong robustness.

**Table 2 Robustness tests**

| Attack | Pop music | | Country music | | Blues music | |
|---|---|---|---|---|---|---|
| | Detected times | BER | Detected times | BER | Detected times | BER |
| No attack | 8 | 0 | 8 | 0 | 10 | 0 |
| Additive noise (55dB) | 7 | 0 | 8 | 0 | 10 | 0 |
| Additive noise (45dB) | 8 | 0 | 8 | 0 | 8 | 0 |
| Additive noise (35dB) | 3 | 0 | 4 | 0.0078 | 5 | 0.0078 |
| Re-quantization | 8 | 0 | 8 | 0 | 10 | 0 |
| Resampling (22050Hz) | 8 | 0 | 8 | 0 | 10 | 0 |
| Resampling (11025Hz) | 8 | 0 | 8 | 0 | 8 | 0 |
| low-pass filtering (10kHz) | 7 | 0 | 6 | 0 | 10 | 0 |
| low-pass filtering (8kHz) | 7 | 0 | 7 | 0 | 9 | 0 |
| low-pass filtering (6kHz) | 7 | 0 | 6 | 0 | 10 | 0 |
| low-pass filtering (4kHz) | 6 | 0 | 5 | 0 | 7 | 0 |
| low-pass filtering (2kHz) | 5 | 0 | 3 | 0 | 7 | 0 |
| Random cutting | 7 | 0 | 6 | 0 | 9 | 0 |
| MP3(128kbps) | 7 | 0 | 5 | 0 | 8 | 0 |
| MP3(96kbps) | 7 | 0 | 5 | 0 | 7 | 0 |
| MP3(80kbps) | 5 | 0 | 2 | 0.0156 | 5 | 0 |
| MP3(64kbps) | 4 | 0 | 2 | 0.0156 | 6 | 0.0156 |

To evaluate the performance of the proposed algorithm, we compare with literature [10-12, 19] against low-pass filter processing. Our algorithm and comparative literatures all give results for lots of audio clips in experiments. Here, we adopt the average value as the way in literature[11], and indicate different cutoff frequencies, as table 3 shows. As we can see, our algorithm have better performance against low-pass filter processing in the case of close embedding capacity.

Table 3 Performance comparison against low-pass filter

|  | Literature [10] | Literature [11] | Literature [12] | Literature [19] | Proposed scheme |
|---|---|---|---|---|---|
| Capacity (bps) | 45.9 | 43 | 30.7 | 80 | 76 |
| SNR(dB) | 24.4 | 32.5 | 25.7 | 37.4 | 30.9 |
| ODG | - | -0.57 | -0.63 | -0.75 | -0.42 |
| Synchronization code | Yes | Yes | Yes | No | Yes |
| Low-pass BER(%) (cutoff frequency, kHz) | 0(11.025) | 0(44.1) | 11.0（10） | 2.7(6) | 0 (2) |

- : the literature didn't give the experimental data.

## 5. Conclusion

A robust audio watermark algorithm is proposed in the paper. The algorithm combines the low-pass characteristics of moving average and energy concentration characteristics of DCT, embeds watermark by adjusting the largest coefficient of DCT domain. Experimental results show that the watermarked audio have high SNR and good audibility. The algorithm is strongly robust against common signal processing operations such as additive noise, re-quantization, resampling, low pass filtering, random cutting, and MP3 compression.

**Acknowledgments**

This work is supported by the Scientific Research Foundation of CUIT under Grant No. KYTZ201420 and Scientific research project of department of education in Sichuan province under Grant No. 16ZA0221.

## Authors

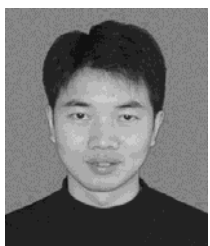

**Jinquan Zhang** received his Ph.D. from Southwest Jiaotong University, Chengdu, China, in 2013. His research areas include audio watermarking, cryptography, digital signature and network security. He is an Assistant Professor in College of Information security Engineering, Chengdu University of Information Technology.

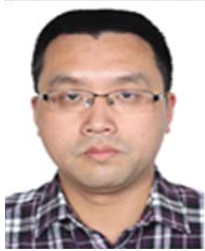

**Bin Han** received the B.S. degree from the Northeast Normal University of China (NENU), Changchun, in 1997 and the M.S. degree from the University of Electronic Science and technology (UESTC), Chengdu, in 2006, both in computer Science. His research interests include information hiding, digital watermarking and network engineering.